\documentclass{aa}  

\usepackage{graphicx, caption}
\usepackage{natbib}
\usepackage{txfonts}
\usepackage{amsmath}
\usepackage{amsmath,siunitx}
\usepackage{multirow}
\usepackage{float}
\usepackage[colorlinks=true,     linkcolor=blue, citecolor=blue, filecolor=blue, urlcolor=blue]{hyperref}
\usepackage{orcidlink}
\usepackage{comment}
\usepackage{setspace}
\begin{document}


   \title{DESI survey of \ion{S}{IV} absorption outflows in quasars: Contribution to AGN feedback and comparison with [\ion{O}{III}] emission outflows}

   \author{Mayank Sharma
          \inst{1}\fnmsep\thanks{\email{mayanksh@vt.edu}} \orcidlink{0009-0001-5990-5790},
          Nahum Arav
          \inst{1}\orcidlink{0000-0003-2991-4618},
          Maryam Dehghanian
          \inst{2}\orcidlink{0000-0002-0964-7500},
          Gwen Walker
          \inst{1}\orcidlink{0000-0001-6421-2449},
          Kyle Johnston
          \inst{1}\orcidlink{0009-0006-4629-7032},\\
          Matthew Kaupin
          \inst{1}\orcidlink{0009-0002-9025-0435}
          \and
          Justin Gandhi
          \inst{1}
          }

   \institute{Department of Physics, Virginia Tech, Blacksburg, VA 24061, USA
   \and
   Department of Physics and Astronomy, University of Kentucky, Lexington, KY 40506, USA}

   \date{}

 
  \abstract
   {}
   {Quasar outflows can play a crucial role in the evolution of their host galaxies through various feedback processes. This effect is expected to be particularly important when the universe was only 2-3 billion years old, during the period known as cosmic noon. By utilizing existing observations from the Dark Energy Spectroscopy Instrument (DESI), we conduct a survey of high-ionization quasar outflows at cosmic noon, with the aim of doubling the current sample of such outflows with distance and energetics determination. We also aim to compare these properties to those derived from spatially resolved outflows in similar quasars probed through Integral Field Spectroscopy (IFS).}  
   {We perform Monte-Carlo simulations on a sample of 130 quasars and detect signatures of high-ionization outflows in the form of \ion{S}{IV} trough in eight objects. The absorption features for each outflow are then analyzed individually to characterize their physical conditions by determining the total hydrogen column density ($N_H$), the ionization parameter ($U_H$), and the electron number density ($n_e$). Through these parameters, we obtained the distance of the outflows from their central source ($R$), their mass outflow rate, and their kinetic luminosity.}
   {The detected outflows show complex kinematic structures with a wide range in blueshifted velocities (100$-$4600 km s$^{-1}$). We locate five out of the eight outflows at distances between 240$-$5500 pc away from the central source. Only upper limits could be obtained for two outflows, placing them closer than 100 and 900 pc; and for one outflow, the distance could not be determined. From the combined sample of 15 high-ionization \ion{S}{IV} outflows at cosmic noon, we find a high fraction (up to 46\%) of them to be powerful enough to contribute significantly to multi-stage AGN feedback processes. Their energetics are also found to be consistent with spatially resolved outflows in a luminosity and redshift matched sample of quasars. Comparison with previous spectra reveals interesting variations in some objects, with two cases of emerging high-velocity broad absorption line features with velocities of $-$8000 and $-$39,000 km s$^{-1}$. An impressive case of four line-locked \ion{Si}{IV} outflow systems is also revealed in one of the objects.}
   {}


    \titlerunning{Quasar Outflows at Cosmic Noon: A DESI View}
   \authorrunning{Mayank Sharma et al.}
   \maketitle
%

\section{Introduction}

Luminous quasars are known to drive powerful outflows that are expected to carry significant mass, momentum and energy away from their central region. These outflows are now observed ubiquitously and in multiple forms: from ultra-fast outflows in X-rays probing the nuclear region \citep[e.g.,][]{reeves2003massive, chartas2021multiphase, matzeu2023supermassive} to ionized, atomic and molecular gas at scales ranging from a few parsecs to kilo-parsecs \citep[e.g.,][]{fiore2017agn, arav2020hst, wylezalek2022first, cresci2023bubbles}. They form the basis of the quasar mode of Active Galactic Nuclei (AGN) feedback which has become fundamental to our current understanding of how supermassive black holes (SMBHs) and their host galaxies co-evolve \citep[e.g.,][]{di2005energy, hopkins2009small}. \par
Observational studies of quasar outflows primarily utilize one of the following two techniques: (a) Integral Field Spectroscopy (IFS) along with long-slit spectroscopy, which spatially resolves the outflowing gas in the plane of sky the and (b) Absorption spectroscopy, which studies the outflowing gas moving towards us, along our line-of-sight to the quasar. These techniques have been applied with both space-based and ground based telescopes to study outflows from the local universe to deep into the epoch of re-ionization. Space-based observations (with telescopes such as Hubble Space Telescope (HST), James Webb Space Telescope (JWST) and Chandra), while being relatively sparse, have provided unprecedented resolution and sensitivity that have allowed to characterize different properties of quasar outflows in great depth. On the other hand, ground based observations (with telescopes such as Very Large Telescope (VLT), Keck and Atacama Large Millimeter Array (ALMA)) have led to large and diverse samples, allowing statistical studies of outflow properties and their contribution to AGN feedback. \par
Based on both theoretical models and observations, quasar-mode of feedback is expected to be the most prevalent during cosmic noon ($z$ $\sim$ 2$-$3), a period marked by intense star formation and black-hole activity \citep[e.g.,][]{boyle1998cosmological, madau2014cosmic, rigby2015cosmic}. At these redshifts, current IFS analysis are limited by their spatial resolution as even the best instruments (e.g., JWST) cannot probe regions closer than \ang{;;0.1} ($\sim$ 800 pc) to the central source.  We therefore rely mainly on absorption line analyses to study the physical properties of the outflowing gas on sub-kpc scales. \par 
To assess the contribution of these outflows to various feedback processes, it is important to constrain their distance from the central source ($R$). This is typically done by first obtaining the electron number density ($n_e$), based on the column density ratio of the (collisionally) excited state to the ground state. Combined with photoionization models of the ionization state of the outflow, it leads to determination of $R$ \citep[e.g.,][]{hamann2001high, de2001keck, moe2009quasar, bautista2010distance, arav2013quasar, lucy2014tracing, finn2014compact, xu2018mini,leighly2018z, arav2020hst, choi2022physical, byun2022vlt, balashev2023low, dehghanian2024narrow, dehghanian2025}. In most cases, these distance determination are based on low-ionization species such as \ion{Si}{II} and \ion{Fe}{II}. This is primarily because for $z \gtrsim 1$ these species have multiple transitions (both ground and excited state) that (a) fall in the observed wavelength range accessible by large ground-based surveys and (b) are long-ward of the rest-frame \ion{Ly}{$\alpha$} transition and thus avoid the contamination posed by the \ion{Ly}{$\alpha$} forest. Most outflows, however, only show absorption troughs from high-ionization species such as \ion{C}{IV}, \ion{N}{V} and \ion{O}{VI} \citep{dai2012intrinsic} and therefore their characteristics might not be well represented by the results derived mainly from lower ionization species. The study of high-ionization outflows from the ground, in most cases, are based on the \ion{S}{IV} resonance and excited state transitions at 1062 \r{A} and 1073 \r{A}, respectively. While other distance indicators such as \ion{N}{III} and \ion{O}{IV} exist at shorter wavelengths, their coverage from ground-based observations is limited to objects with $z \gtrsim 3$ where the increasing density of the \ion{Ly}{$\alpha$} forest makes it difficult to obtain reliable trough measurements. \cite{xu2019vlt} presented results from a dedicated VLT/X-Shooter survey of 8 quasar outflows at $z \sim 2$, in which \ion{S}{IV} was detected and obtained the distance and energetics for the sample. \cite{arav2018evidence} analyzed a much larger sample of 34 quasar outflows from the SDSS, constraining their $R$ and showing that most broad absorption line (BAL, with width $\Delta v >$ 2000 km s$^{-1}$) outflows have $R \gtrsim$ 100 pc. In this paper, we build upon these works using the Early Data Release (EDR) of the Dark Energy Spectroscopy Instrument \citep[DESI, see][for a detailed description of the instruments and the data release.]{abareshi2022overview, adame2023early}. \par
The paper is structured as follows. Section \ref{sec:sample} describes how our parent sample  was constructed from the DESI-EDR. In section \ref{sec:spectral}, we summarize the methodology used for our spectral analysis and modeling (more detailed explanation of the methodology is provided in the appendices). We summarize the results from this analysis in section \ref{sec:results} and also discuss the combined \ion{S}{IV} sample \citep[including objects from][]{xu2019vlt}. In section \ref{sec:IFS}, we compare the properties of the \ion{S}{IV} absorption outflows with the [\ion{O}{III}] emission outflows studied through IFS. Interesting features of individual objects are discussed in section \ref{sec:notes}. Finally, we discuss some important aspects of our results in section \ref{sec:discussion} and conclude with a summary in section \ref{sec:summary}.  Throughout the paper, we adopt a standard $\Lambda$CDM cosmology with h = 0.677, $\Omega_m$ = 0.310 and $\Omega_{\Lambda}$ = 0.690 \citep{planck2018}. 

\section{Sample Selection} \label{sec:sample}

DESI is conducting a spectroscopic survey of the universe with the aim to produce detailed map of its three dimensional structure for which observations began in December 2020. This 5 year project is expected to observe more than 30 million galaxies (including 3 million quasars) right from the local universe to beyond $z \sim 3.5$. Spectra obtained during the 5 month survey validation phase conducted in 2021 (prior to the main survey) were made available through the Early Data Release (EDR) which includes 76,079 quasars. EDR was accompanied by the release of a set of  “value-added catalogs” (VACs) generated by the DESI science collaboration, which includes a BAL quasar catalog \citep[see][for a description of the catalog]{filbert2024broad}. We target this catalog (from the survey validation 1 (sv1) phase) to obtain a suitable sample for our study. \par
The DESI spectrographs cover a wavelength range between 3600 and 9824 \r{A} with spectral resolution increasing from $\sim$ 2000 at 3600 \r{A} to $\sim$ 5500 at 9800 \r{A}. In order to detect the \ion{S}{IV} 1063 \r{A} trough, we restrict our sample to quasars with redshifts $>$ 2.5. Moreover, we only consider objects with a g-band magnitude $\lesssim$ 21.4 to ensure sufficient signal-to-noise (S/N) for our analysis \footnote{As different objects in the DESI EDR are observed for varying exposure times, a smaller magnitude does not always imply higher S/N. However, through visual inspection, we found this cutoff to be sufficient for our analysis.}. This resulted in a sample of $\sim$ 1600 quasars. We visually inspected the spectral region \citep[following the procedure outlined in][]{arav2018evidence} between the \ion{C}{IV} and \ion{Si}{IV} emission lines to identify signature of outflowing gas in the form of relatively-broad (FWHM $\gtrsim$ 500 km$^{-1}$) and moderately-deep (minimum residual intensity (I) $\lesssim$ 0.7) troughs. We then identify outflowing systems that showed absorption troughs from both \ion{C}{IV} and \ion{Si}{IV} resulting in our parent sample of 130 objects. 

\begin{table}
\setlength{\tabcolsep}{6pt}
\renewcommand{\arraystretch}{1.33}
\caption{Properties of the objects in our final sample}
\centering
\begin{tabular}{lccc}
\hline\hline
Object & $z$ & log($L_{Bol}$) & \ion{C}{IV} Abs. Width \\  
 &  & (erg $s^{-1}$) & (km $s^{-1}$)\\  
\hline
J0542-2003 & $3.151^{*}$ & 47.1 & 1700 \\
J0845+2127 & $2.576^{\star}$ & 47.0 & 4100 \\
J0857+3149 & 2.901 & 46.7 & 9000 \\
J1021+3209 & $2.635^{*}$ & 46.6 & 1100 \\
J1035+3253 & 2.694 & 46.1 & 10600 \\
J1054+3158 & $2.676^{\star}$ & 46.8 & 1300 \\
J1229+3316 & 3.032 & 47.0 & 1100 \\
J1426+3151 & $2.792^{\star}$ & 46.5 & 6600 \\
\hline
\end{tabular}
\tablefoot{The redshifts marked with a * or $\star$ were found to be incorrectly determined by the DESI pipeline due to misidentification of emission features. In these cases, the reported redshifts were either adopted from SDSS \citep[][marked with a $\star$]{paris2018sloan} or calculated independently by fitting the \ion{Si}{IV} emission line (marked with a *).}
\label{table:objectpar}
\end{table}

\section{Methodology} \label{sec:spectral}

Analyzing the absorption troughs requires modeling the unabsorbed spectrum of the quasar, which is made up of two components: (a) an underlying continuum and (b) emission lines. We model all the objects in our sample following \cite{sharma2025distance} and find that the continuum is well modeled with a single power law. Prominent emission features such as \ion{Ly}{$\alpha$}, \ion{N}{V} and \ion{C}{IV} require separate broad emission line (BEL) and narrow emission line (NEL) components, whereas \ion{Si}{IV} and \ion{O}{VI} are modeled with a single component. The combined unabsorbed model is then used to obtain the normalized spectra. We summarize the important steps of our analysis below and provide the technical details for the same in Appendces \ref{sec:AppendixA} and \ref{sec:AppendixB}.

\begin{enumerate}
    \item \textit{Identifying \ion{S}{IV} troughs:} Following \cite{arav2018evidence}, we use the kinematic profile of the \ion{Si}{IV} 1403 \r{A} trough to confirm the detection of the \ion{S}{IV} 1063 \r{A} trough at its expected wavelength. Based on Monte-Carlo simulations, we identify 8 objects (see Table \ref{table:objectpar}) that show the \ion{S}{IV} trough with $\geq 95\%$ confidence level, making up our final sample. 
    \item \textit{Column Density Determination:} In all eight outflows, we detect troughs corresponding to ions from multiple elements, arising from varying levels of ionization and excitation. We model these troughs with a Gaussian template in velocity space with fixed widths and centroids for each outflow individually. The ionic column densities ($N_{ion}$) are then obtained using these profiles and are reported in Table \ref{table:colden}. 
    \item \textit{Photoionization Modeling:} The measured $N_{ion}$ are a result of the ionization structure of the outflowing gas and can thus be used to constrain its physical state. The absorbing gas is characterized through two important parameters: the total hydrogen column density ($N_H$) and the ionization parameter ($U_H$), which is related to the rate of ionizing photons emitted by the source ($Q_H$) as
    \begin{equation} \label{uh}
    U_H = \frac{Q_H}{4\pi R^2 n_H c}
    \end{equation} 
    Here $R$ is the distance between the emission source and the observed outflow component, $n_H$ is the hydrogen number density and $c$ is the speed of light. Using the photoionization code \textsc{Cloudy} \citep[vers. 23.01,][]{gunasekera202323} and following the procedure described in \cite{edmonds2011galactic}, we compare the measured $N_{ion}$ values with model predictions to obtain best-fit values for $N_H$ and $U_H$.
    \item \textit{Collisional Excitation Modeling:} In seven out of eight outflows in our sample, we detect troughs arising from collisionally excited levels of one or more of the following species: \ion{C}{II}, \ion{Si}{II}, \ion{N}{III} and \ion{S}{IV}. The relative population of these excited states with respect to the ground state is an indicator of the $n_e$ of the outflow \citep[e.g.,][]{arav2018evidence}. Using the \textsc{Chianti} atomic database \citep[vers. 9.0;][]{dere2019chianti}, we predict the theoretical population ratios as a function of $n_e$. By comparing them with the measured column density ratios, we can obtain the corresponding $n_e$ for each outflow.
\end{enumerate}

\section{Results} \label{sec:results}

\subsection{Physical Properties of the Outflowing Gas}

\begin{figure}
   \centering
   \resizebox{\hsize}{!}
        {\includegraphics[width=1.00\linewidth]{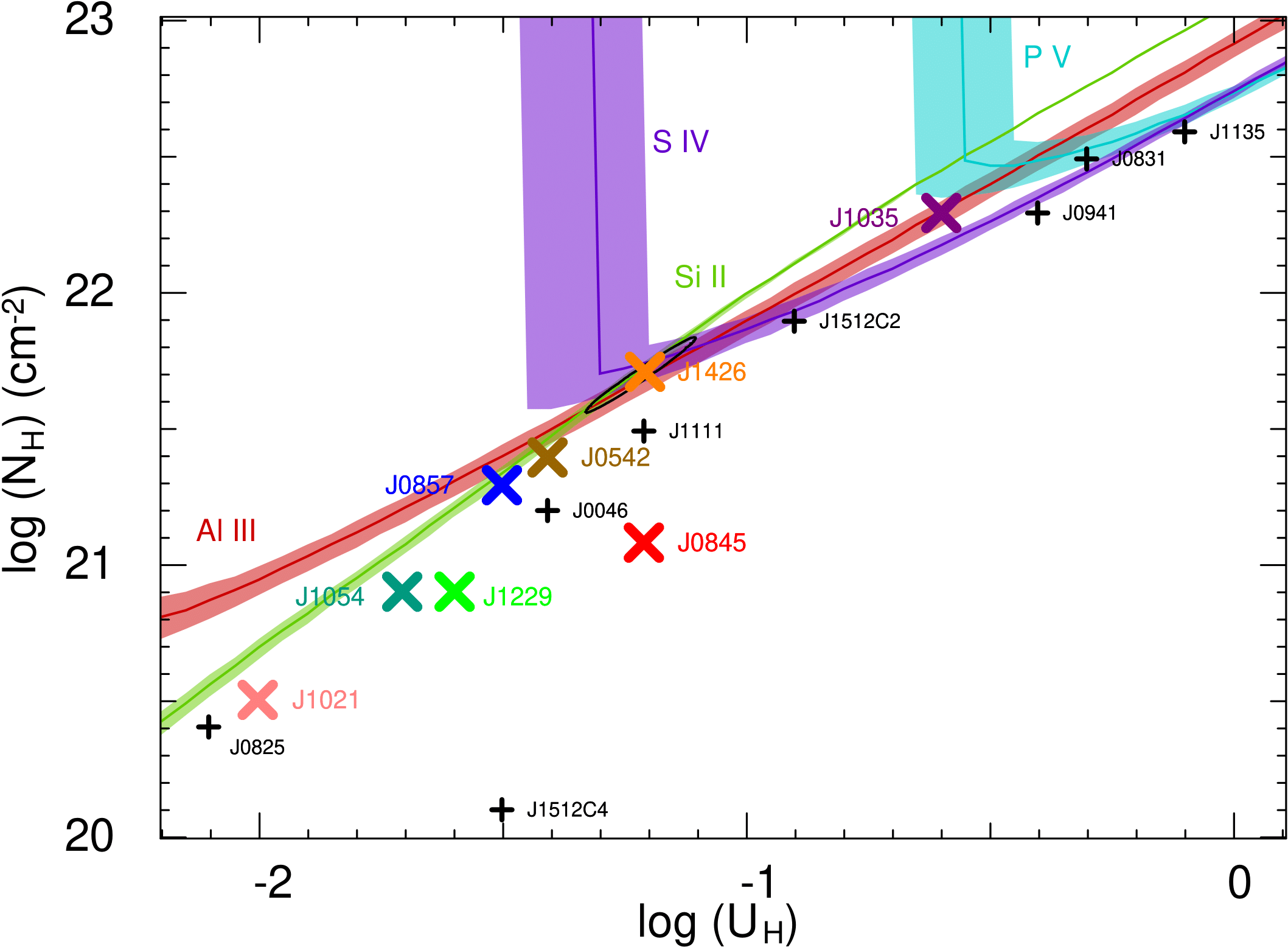}}
      \caption{Phase-space plot showing the $N_H$ and $U_H$ solutions for the DESI sample (colored X) and the VLT sample of \cite{xu2019vlt} (black plus signs).  The colored lines and associated shaded regions show the  $N_{ion}$ constraints (measurements and errors, respectively) used to derive the   $N_H$ and $U_H$ solutions for the outflow detected in J1426+3151. The phase-space solution with minimized $\chi^2$ is shown as an orange cross surrounded by a black eclipse indicating the 1$\sigma$ error.}
         \label{fig:nvu}
\end{figure}

Fig. \ref{fig:nvu} shows the best-fit parameters in the ($N_H$, $U_H$) phase-space derived from the photoionization modeling of the outflows. The solutions spread over 2 dex in log($N_H$) and 1.5 dex in log($U_H$), and lie close to or before the hydrogen ionization front (traced roughly by the line log($N_H$) - log($U_H$) $\sim$ 23), consistent with the expectations for high ionization outflows \citep{lucy2014tracing}. Both these features are also similar to the results from the VLT sample of \cite{xu2019vlt} (shown with black plus signs). 

\begin{figure}[]
         \centering
         \includegraphics[width=1.00\linewidth]{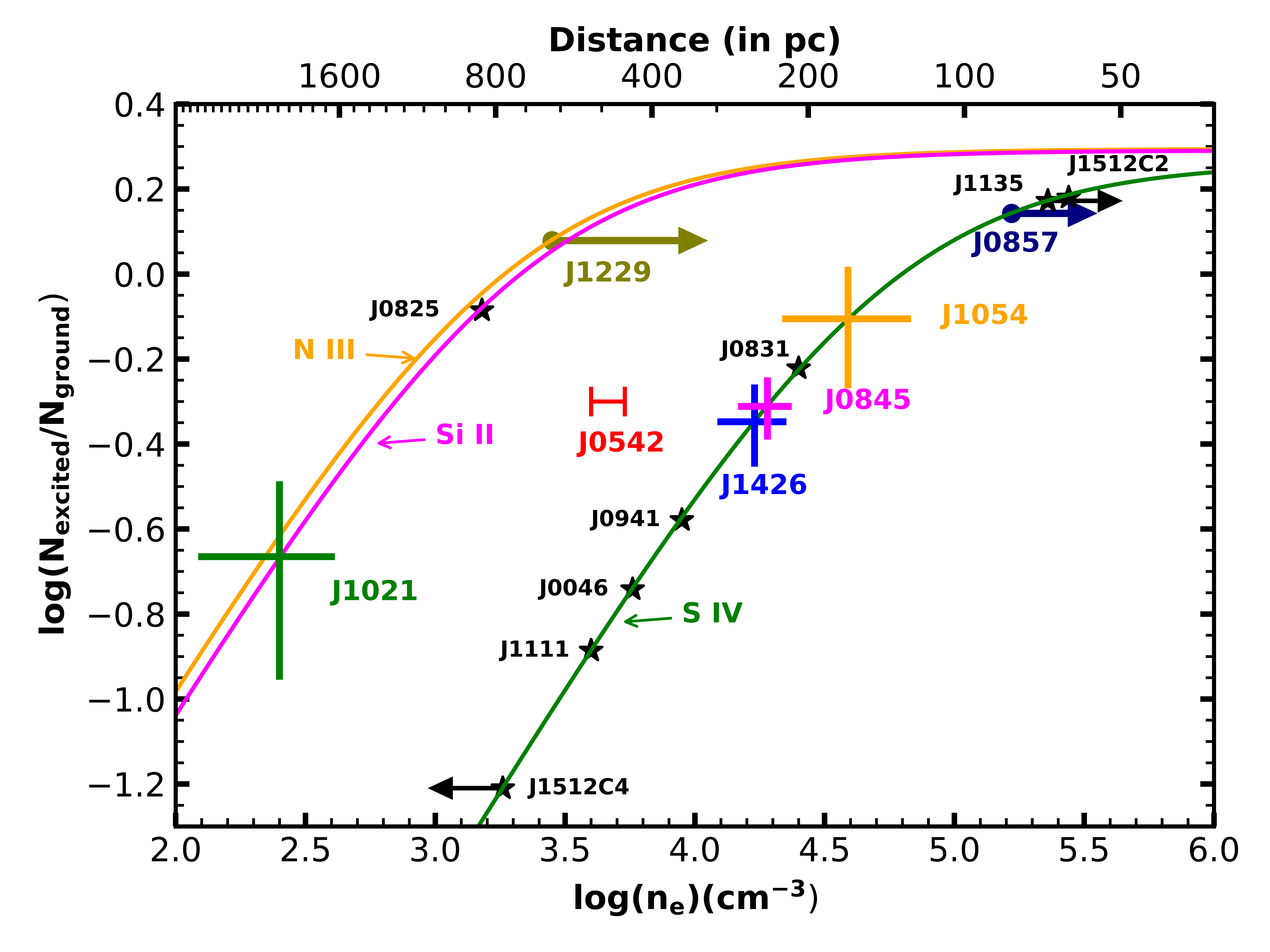}
         \caption{Theoretical population ratios for different ionic species for $T_e$ = 15,000 K (solid curves). The vertical part of the crosses represent the observed column density ratios and their error for different outflowing systems. The horizontal part of the crosses represent the associated $n_e$ error. The arrows mark the lower limits obtained from possibly saturated troughs corresponding to these species (The case of J0542 is explained in section \ref{sec:0542}). The top axis shows the distance scale based on the outflow detected in J1426+3151. The black stars (and arrows) correspond to the objects from \cite{xu2019vlt}.}
         \label{fig:chianti}
\end{figure}

In the case of J0542-2003 and J0857+3149, we obtained robust column density measurements for two ionic species from the same element: \ion{Al}{II} and \ion{Al}{III}. Their ratio allows us to obtain the ionization state of the outflow independent of the assumed abundances. In both cases, we find that the ($N_{H}, U_{H}$) solution obtained using all measured ionic column density constraints (and assumed solar abundance) does not differ significantly compared to the abundance independent solution obtained using only \ion{Al}{II} and \ion{Al}{III} (with $\Delta$ $\sim$ 0.1 dex for both parameters). This indicates that the actual gas-phase abundance in these outflows is indeed close to our initial assumption of solar values \citep[based on ][which is the default used by \textsc{Cloudy}]{grevesse1998standard}.\\
Fig. \ref{fig:chianti} shows the $n_e$ obtained from the collisional excitation modeling of the outflows. By utilizing excited states corresponding to three different ionic species over a broad range of ionization, we are able to obtain log($n_e$) solutions spreading over 3 dex, which is similar to the range probed by \cite{xu2019vlt} in their VLT sample. For some of the outflowing systems, we measured column densities from multiple ground and excited state transitions. In all these cases the limits/measurements obtained from all the excited states were found to be consistent and we only show the most robust determination. \par
The $n_e$ solutions in Fig. \ref{fig:chianti} were obtained for an assumed temperature of 15,000 K. In principle, as the outflows have different $U_H$, they can have different effective temperatures. To account for this effect, we obtain the effective temperature in each cloud for all the ionic species used in the $n_e$ determination. We find that they span a range of temperatures between 11,000 K and 17,000 K. Including this effect changes the obtained log($n_e$) solutions by less than 0.1 dex.

\subsection{Distance and Energetics}

Having determined $U_H$ and $n_e$, we can use equation (\ref{uh}) to obtain the distance of the outflowing components from their central source ($R$) as:

\begin{equation} \label{distance}
    R = \sqrt{\frac{Q_H}{4\pi U_H n_H c}}.
    \end{equation}

For highly ionized plasma, $n_H$ and $n_e$ are related as $n_H \approx$ $0.8 n_e$ \citep{osterbrock2006astrophysics}. To calculate $Q_H$, we scale the spectral energy distribution (SED) to match the observed flux density ($F_{\lambda}$) at rest wavelength $\lambda_{rest}$ $\approx$ 2900 \r{A} for each object and then integrate the scaled SEDs for energies above 1 Ryd. We report the determined distance for the outflowing components in Table \ref{table:outflowpar}. We find that five out of the seven outflows are located at distances larger than 200 pc, including two outflows with $R$ > 1 kpc. For the outflows in J0857+3149 and J1229+3316, we only have upper limits on $R$ that places them closer than 100 pc and 900 pc to the central source respectively. \par
The mass outflow rate ($\dot{M}$) and kinetic luminosity ($\dot{E}_k$) are given by \citep[see][for details]{borguet201210}
\begin{equation}
\dot{M}\simeq 4\pi \Omega R N_H \mu m_p v \ \ \ \mbox{and} \ \ \   \dot{E}_k= \frac{1}{2} \dot{M} v^2, 
\label{eqn:energetics}
 \end{equation}
where $m_p$ is the proton mass, $\mu \approx 1.4$ is the mean atomic mass of the gas per proton and $\Omega$ is the fractional solid angle covered by the outflow. Typically, the frequency of detection of an outflow type in all quasar spectra is used as an estimate for $\Omega$. For BAL outflows, the incidence rate is $\sim$ 20\%, based on the detection of \ion{C}{IV} \citep{hewett2003frequency, knigge2008intrinsic}. As \cite{dunn2012bal} showed using photoionization models, absorption troughs from \ion{S}{IV} and \ion{C}{IV} are likely to arise from the same physical region for a given outflow and thus have the same $\Omega$. Therefore, we use $\Omega$ = 0.2 to obtain $\dot{M}$ and $\dot{E}_k$ for the outflows in our sample and report them in Table \ref{table:outflowpar}. While obtaining their errors, we account for the correlation between $R$ and $N_H$ (through the error ellipse in Fig. \ref{fig:nvu}) following the procedure described in \cite{walker2022high}. 

\begin{table*}
\setlength{\tabcolsep}{6pt}
\renewcommand{\arraystretch}{1.33}
\caption{Outflow Parameters}
\centering
\begin{tabular}{lcccccccc}
\hline\hline
Object & $v$ & log($U_H$) &log($N_H$) &log($n_e$) & $R$ & $\dot{M}$ & log($\dot{E}_k$) & $\dot{E}_k/L_{Bol}$\\
    &  (km $s^{-1}$)   & dex       & log(cm$^{-2}$)& log(cm$^{-3}$) & pc & ($M_{\odot}$ yr$^{-1}$) & log(ergs $s^{-1}$)  &\%\\  
\hline
J0542-2003 & -2520 & $-1.4^{+0.1}_{-0.2}$  & $21.4^{+0.2}_{-0.2}$  & $3.72^{+0.07}_{-0.06}$  &$1100^{+250}_{-200}$ & $198^{+48}_{-45}$  & $44.60^{+0.09}_{-0.11}$ & $0.37^{+0.08}_{-0.08}$   \\ 
J0845+2127 & -3800 & $-1.2^{+0.5}_{-0.3}$ & $21.1^{+0.6}_{-0.3}$  & $4.33^{+0.09}_{-0.09}$  & $400^{+160}_{-190}$   & $53^{+56}_{-17}$  & $44.39^{+0.31}_{-0.16}$ & $0.22^{+0.23}_{-0.07}$\\
J0857+3149 & -4580 & $-1.5^{+0.1}_{-0.2}$  & $21.3^{+0.2}_{-0.2}$  & $>5.55_{-0.3}$ & $<100^{+50}$  & $<29^{+14}$   &$<44.28^{+0.18}$ & $<0.38^{+0.19}$\\
J1021+3209 & -100 & $-2.0^{+0.1}_{-0.1}$ & $20.5^{+0.1}_{-0.2}$  & $2.46^{+0.19}_{-0.31}$  & $5500^{+2500}_{-1200}$  & $5^{+2}_{-1}$ & $40.15^{+0.18}_{-0.15}$& $3.7^{+1.9}_{-1.0}$ $\times 10^{-5}$\\
J1035+3253 & -2570 & $-0.6^{+0.3}_{-0.3}$ & $22.3^{+0.4}_{-0.3}$  & - & - & - & - & -\\
J1054+3158 & -1390 & $-1.7^{+0.5}_{-0.3}$  & $20.9^{+0.5}_{-0.3}$  & $4.65^{+0.22}_{-0.25}$  & $390^{+190}_{-180}$   & $13^{+12}_{-4}$  & $42.90^{+0.29}_{-0.17}$  & $0.014^{+0.013}_{-0.004}$ \\
J1229+3316 & -2660 & $-1.6^{+0.4}_{-0.2}$  & $20.9^{+0.6}_{-0.3}$  & $>4.05_{-0.57}$ &  $<900^{+900}$  & $<52^{+94}$   &$<44.07^{+0.45}$ & $<0.1^{+0.2}$\\
J1426+3151 & -2330 & $-1.2^{+0.1}_{-0.1}$  & $21.7^{+0.1}_{-0.2}$  & $4.29^{+0.10}_{-0.13}$ & $240^{+60}_{-40}$   & $81^{+21}_{-17}$  & $44.14^{+0.10}_{-0.10}$  & $0.49^{+0.13}_{-0.11}$\\
\hline
\end{tabular}
\label{table:outflowpar}
\end{table*}

\subsection{The combined S IV sample}

Combining the \ion{S}{IV} objects from \cite{xu2019vlt} with our sample leads to a total of 15 quasars in the redshift range 2.00 $\leq z \leq$ 3.15. These contain 16 analyzed \ion{S}{IV} outflows (J1512+1119 has two outflows), out of which 10 have robust distance (and energetics) measurements, 5 have upper/lower limits and for one no constraints could be obtained. From this, we exclude two objects from \cite{xu2019vlt} (i) J0825+0740: as it has an outflow with positive velocity, indicating either an infall or uncertainty in the quasar redshift determination and (ii) J1512+1119: as the $R$ for its outflow is only constrained to within a factor of 30. \par

\begin{figure}[]
         \centering
         \includegraphics[width=1.00\linewidth]{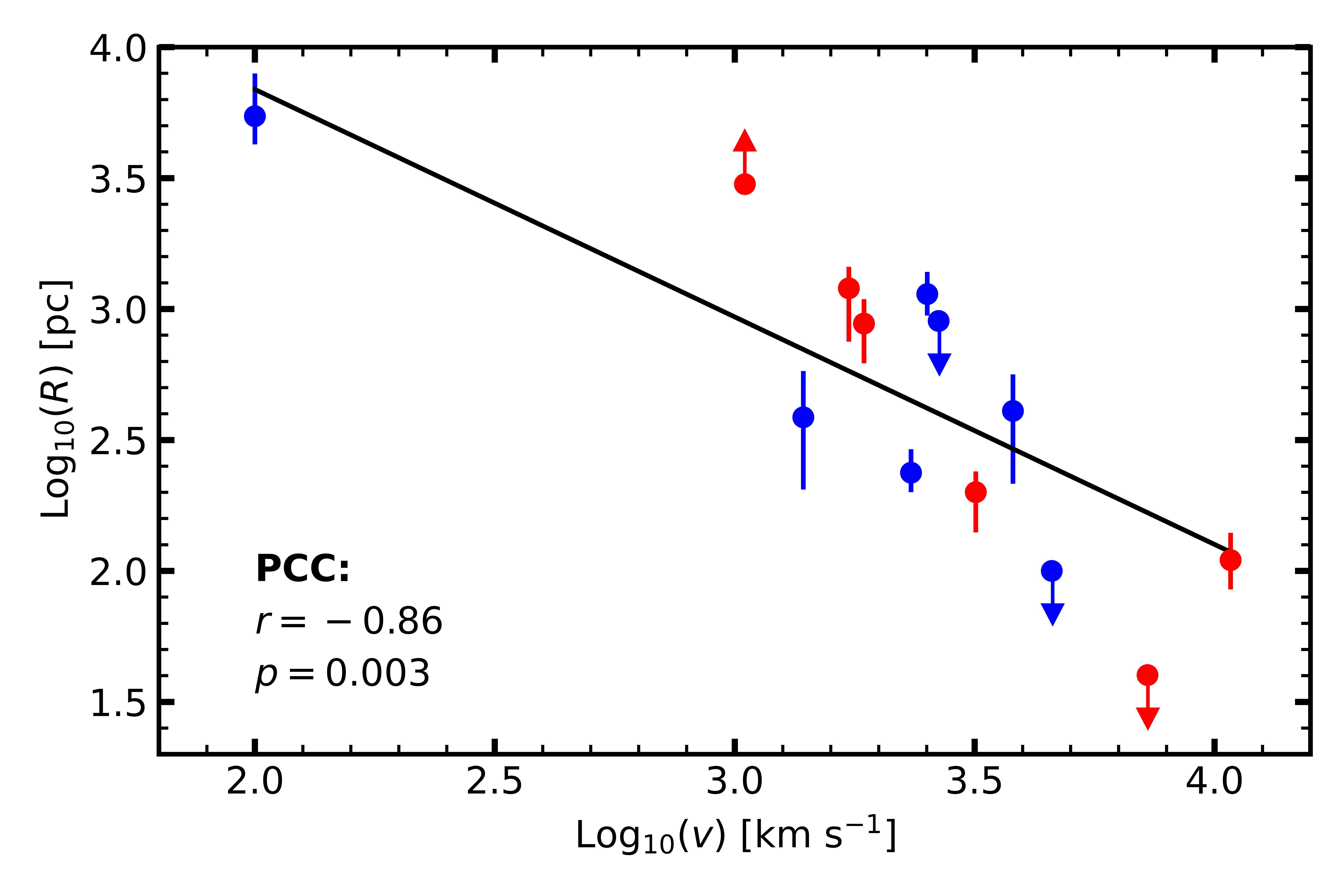}
          \caption{Distribution of log($R$) vs log($|v|$) for the combined \ion{S}{IV} sample. The objects analyzed in this work are shown in blue, and the objects from \cite{xu2019vlt} are shown in red. The sample shows strong negative correlation with a Pearson's correlation coefficient (PCC) of $-$0.86 (which drops to $-$0.68 if the lowest velocity outflow is excluded). The black line shows the best-fit log-linear relationship given by log($R$) = $-$0.87 log($|v|$) + 5.58.}
         \label{fig:RvsV}
\end{figure}

In Fig. \ref{fig:RvsV}, we plot the distance of the outflows as a function of their line-of-sight velocities which reveals that faster outflows are typically found to be closer to their central source. For this trend, the Pearson's correlation coefficient (PCC) is found to be $r =-$ 0.86 thus suggesting a strong negative correlation. We find the p-value to be $p = 3\times10^{-3}$, indicating only a $\sim$ 0.3\% probability that a normal uncorrelated system could produce the observed trend. We only have one data point in the low-velocity regime where the errors in velocity determination are affected greatly by the accuracy of redshift determination. To ensure that this data point alone does not drive our correlation statistics, we remove it from our sample and recalculate $r$ and $p$. This leads to $r =-$ 0.68 (with $p = 6.3\times10^{-2}$) still indicating moderate to strong negative correlation. We perform a $\chi^2$ minimization to find the best fit log-linear relationship (shown as the black solid line in Fig. \ref{fig:RvsV}) with slope of $-$0.87 and a y-intercept of 5.58. These values are similar to the ones obtained by \cite{byun2024bal} ($-$1.08 and 6.44, respectively) for a sample of 24 very-high ionization absorption outflows detected in the extreme UV using HST. These observed trends are possibly a result of the outflowing gas slowing down naturally as it moves farther away, sweeping up more material and is expected in theoretical models of energy-conserving shock bubbles \citep[see Fig. 1 and 3 of][]{hall2024quasar}. \par
Next, we compare the velocity and mass outflow rate of the outflows with the quasar luminosity ($L_{\textrm{Bol}}$). While more luminous quasars do tend to drive faster and more powerful outflows in general, we find the correlation of both $v$ and $\dot{M}$ with $L_{Bol}$ to be weak ($r \sim$ 0.42 and $p \sim$ 0.25 in both cases). This could be a result of the small range ($\leq$ 1 dex) in luminosities spanned by our sample and also due to the fact that the observed luminosities could be different from their long-term averages.

\section{Comparison with IFS Analyses} \label{sec:IFS}

Multiple IFS analyses have revealed large-scale spatially resolved outflows in luminous quasars. To compare the properties of these outflows with our absorption analysis, we focus on two studies that probe quasars with similar average luminosity ($\sim$ $10^{47}$ erg s$^{-1}$) and redshift ($\sim$ 2.5) as that of our combined \ion{S}{IV} sample: \cite{carniani2015ionised} and \cite{kakkad2020super}. The outflows in both these studies are primarily traced by the $\lambda_{rest}$ = 5007 \r{A} emission feature of [\ion{O}{III}], an ion with ionization potential of $\sim$ 55 eV. This is similar to some of the important ionic species detected in our absorption sample (e.g., 47 eV for \ion{S}{IV} and 64 eV for \ion{C}{IV}) and thus these two manifestations are expected to trace outflowing gas with similar physical conditions \citep[][]{crenshaw2005connection, xu2020evidence}. \par
\begin{figure}[]
         \centering
         \includegraphics[width=1.00\linewidth]{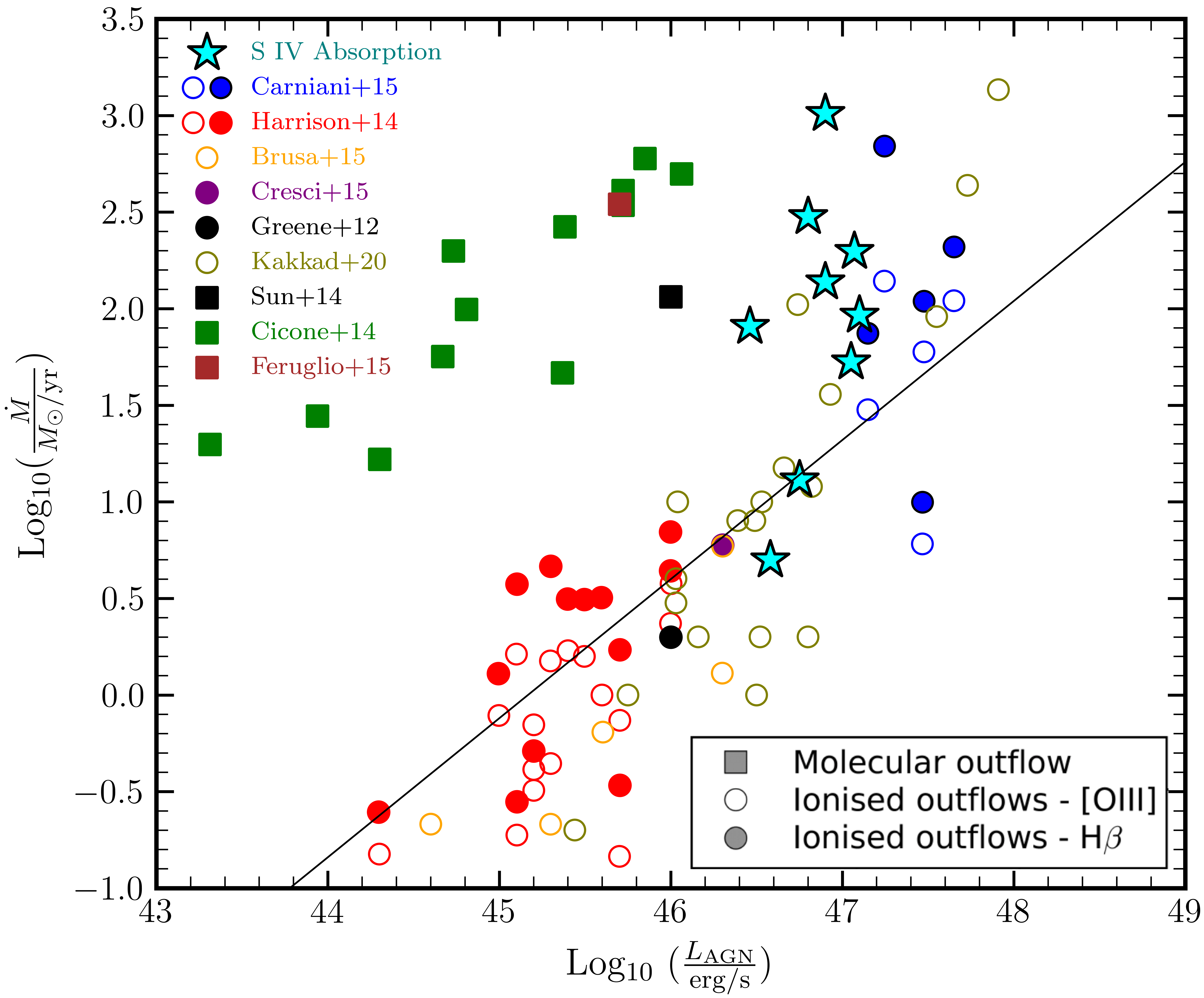}
         \caption{Mass outflow rates ($\dot{M}$) as a function of the AGN bolometric luminosity \citep[adapted from Fig. 9 of][with permission]{carniani2015ionised}. The teal stars represent our combined \ion{S}{IV} sample seen in absorption. The open and filled circles represent the $\dot{M}$ for ionized outflows determined from [\ion{O}{III}] and \ion{H}{$\beta$} emission lines, respectively. The squares denote molecular outflows.}
         \label{fig:MassOutflowCarniani}
\end{figure}

In Fig. \ref{fig:MassOutflowCarniani} we show the relation between the $\dot{M}$ of outflows and the bolometric luminosity (referred to as $L_{AGN}$) of their central quasar. The samples of \cite{carniani2015ionised} and \cite{kakkad2020super} are shown as blue and olive circles respectively. We find that the absorption outflows (shown as teal stars) occupy a similar region in the $\dot{M}-L_{\textrm{AGN}}$ phase-space as the spatially resolved ionized outflows. A similar trend is found while comparing the outflow-momentum  ($\dot{M}v$) to the photon momentum of the AGN ($L_{\textrm{AGN}}/c$) \cite[Fig. 11 of][]{carniani2015ionised}, with the absorption outflows covering a wider range in outflow-momentum due to their larger velocity range. The similarity between the characteristics obtained from independent analysis for the two (luminosity and redshift matched) samples provides further evidence for the connection between the emission and absorption outflows and their common physical origin \citep{sharma2025distance, zhao2025galactic}. \par
Fig. \ref{fig:MassOutflowCarniani} also shows outflows from ionized and molecular outflows in other samples of lower luminosity AGN \citep[from][]{harrison2014kiloparsec, brusa2015x, cresci2015blowin, greene2012spectacular, sun2014alma, cicone2014massive, feruglio2015multi}. The black line represents the scaled best-fit log-linear relationship obtained for ionized outflows \citep[by][]{carniani2015ionised} by fixing the slope to the relationship obtained for molecular outflows by \cite{cicone2014massive}. As only a small sample of ionized outflows in luminous quasars were analyzed up to this point, the intercept was mainly constrained by the low luminosity regime. Including the sample of \cite{kakkad2020super} shows that for ultra-luminous quasars ($L_{AGN}\gtrsim$ $10^{47}$ erg s$^{-1}$ ) the obtained mass outflow rate is consistently underestimated by the best-fit relationship. This is supported by the results from the absorption analysis and indicates a steeper slope for ionized outflows as compared to the molecular outflows. This is consistent with the distinct scaling relations obtained by \cite{fiore2017agn} for molecular ($\dot{M} \propto L_{AGN}^{0.76\pm0.06}$) and ionized outflows ($\dot{M} \propto L_{AGN}^{1.29\pm0.38}$) by considering a wider range in luminosity. The discrepancy between the mass outflow rates of molecular and ionized outflows is also highlighted through Fig. \ref{fig:MassOutflowCarniani}. At low luminosities, the ratio between the molecular and ionized mass outflow rates is $\sim$ 100. Due to the steeper slope for the ionized winds, this ratio is expected to decrease with an increase in luminosity. This trend, however, remains poorly constrained empirically above $L_{AGN} > 10^{46}$ erg s$^{-1}$, as very few molecular outflows have been studied in such objects \citep[e.g.,][]{carniani2017agn, spilker2025direct}. 

\section{Notes on individual objects} \label{sec:notes}

\subsection{J0542-2003: $n_e$ determination and line-locking} \label{sec:0542}
For the outflow analyzed in this object, we obtain log($n_e$) = $3.72^{+0.07}_{-0.06}$. Unlike other objects in our sample, this determination was not based on any single ionic species and therefore the solution did not lie on any of the theoretical population curves in Fig. \ref{fig:chianti}. This is because we obtained a lower-limit based on the \ion{Si}{II*} 1265 \r{A} trough and an upper limit
based on the non-detection of \ion{S}{IV*} 1073 \r{A} trough. These two limits tightly constrained the solution from both ends within 0.13 dex. \par
\begin{figure}[]
         \centering
         \includegraphics[width=1.00\linewidth]{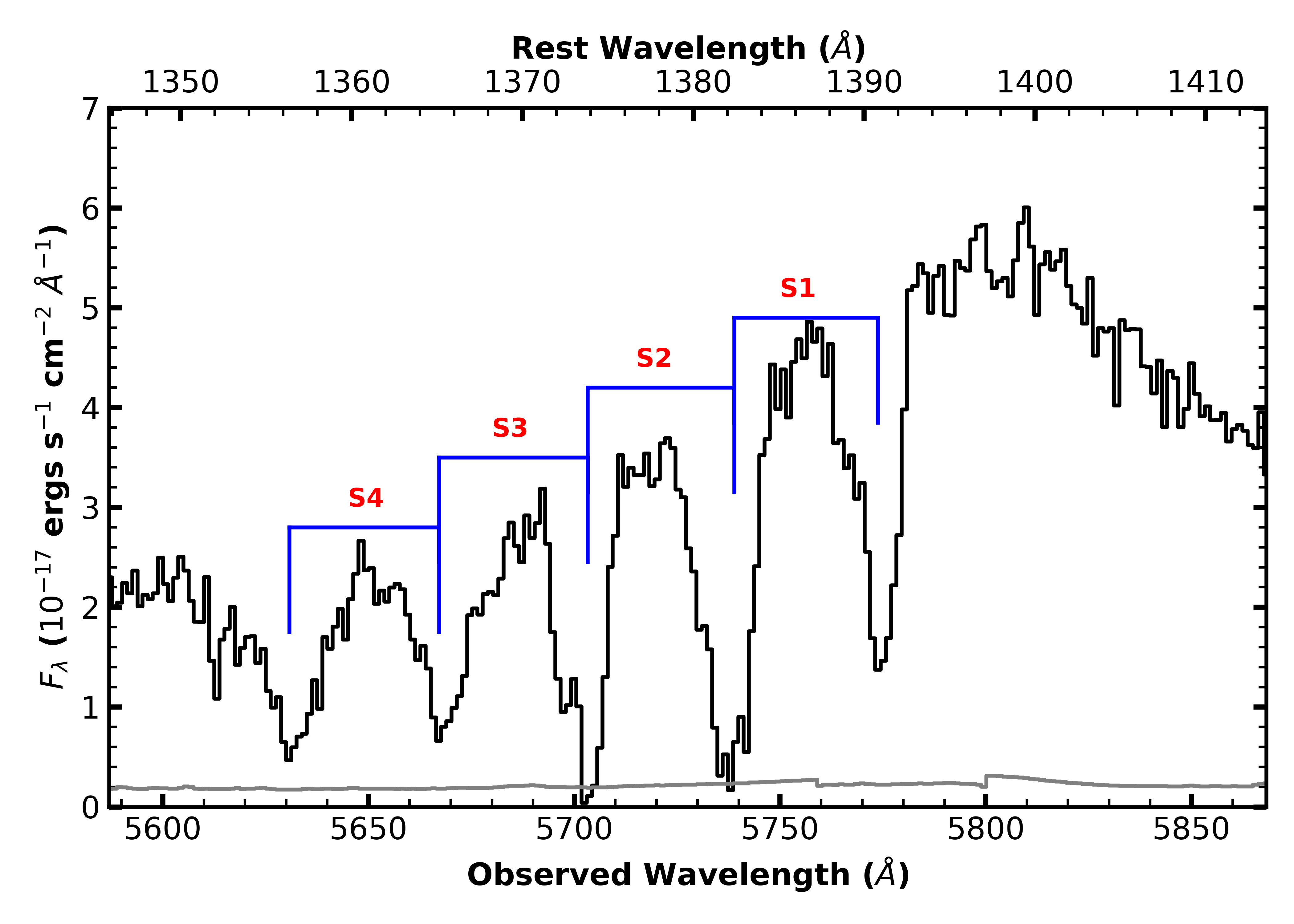}
         \caption{Unnormalized DESI spectrum of J0542-2003 highlighting the multiple line-locked \ion{Si}{IV} doublets. The four absorption systems (S1 through S4) are separated in velocity space by the same amount as the separation between the \ion{Si}{IV} doublets leading to the nested absorption features.}
         \label{fig:linelock}
\end{figure}

Apart from the main outflowing component analyzed in this work (S1 at $-$ 2500 km s$^{-1}$), we also detect three other systems through their \ion{Si}{IV} troughs at $-$4400 (S2), $-$6300 (S3) and $-$8200 km s$^{-1}$ (S4) as shown in Fig. \ref{fig:linelock}. The separation between these distinct kinematic components ($\sim 1900$ km s$^{-1}$) is similar to the velocity separation between the \ion{Si}{IV} doublets ($\approx$ 1934 km s$^{-1}$). This phenomenon is referred to as line-locking and serves as an indicator of the significance of radiation pressure in driving quasar outflows \citep{goldreich1976quasar, braun1989non, lewis2023fine}. The three new components are also detected in other high ionization species such as \ion{C}{IV}, \ion{N}{V} and \ion{O}{VI}, while only S2 shows troughs from low-ionization species such as \ion{C}{II} and \ion{Al}{III}. The presence of excited states corresponding to the additional line-locked systems can neither be confirmed confidently nor ruled out due to the limited resolution and S/N at lower wavelengths. We are thus unable to constrain their distance and energetics at this stage.

\subsection{J0845+2127: Trough variability and BAL emergence}
The spectrum of J0845+2127 also contains additional kinematic components (along with the main outflowing system with $v$ = $-$3800 km s$^{-1}$ (S1)) at $-$6200 (S2) and $-$22000 km s$^{-1}$ (S3) with widths of $\sim$ 1200 km s$^{-1}$ and 3000 km s$^{-1}$ respectively. Both components also show troughs from other high-ionization species such as \ion{Si}{IV}, \ion{N}{V} and \ion{O}{VI}, but none from commonly detected low ionization species. This object was also observed with SDSS on two occasions: November 2012 and January 2018. We compare the two SDSS epochs with the DESI spectrum from January 2021, which reveals interesting changes over $\sim$ 2.3 yrs in the quasar rest frame. We find that the continuum flux level decreased by $\sim$ 30\% between the 2012 and 2018 epochs in the observed wavelength range  3600 - 6000 \r{A}. Between 2018 and 2021, the continuum flux in the same wavelength range increased again by $\sim$ 25\%. \par
\begin{figure}[]
         \centering
         \includegraphics[width=1.00\linewidth]{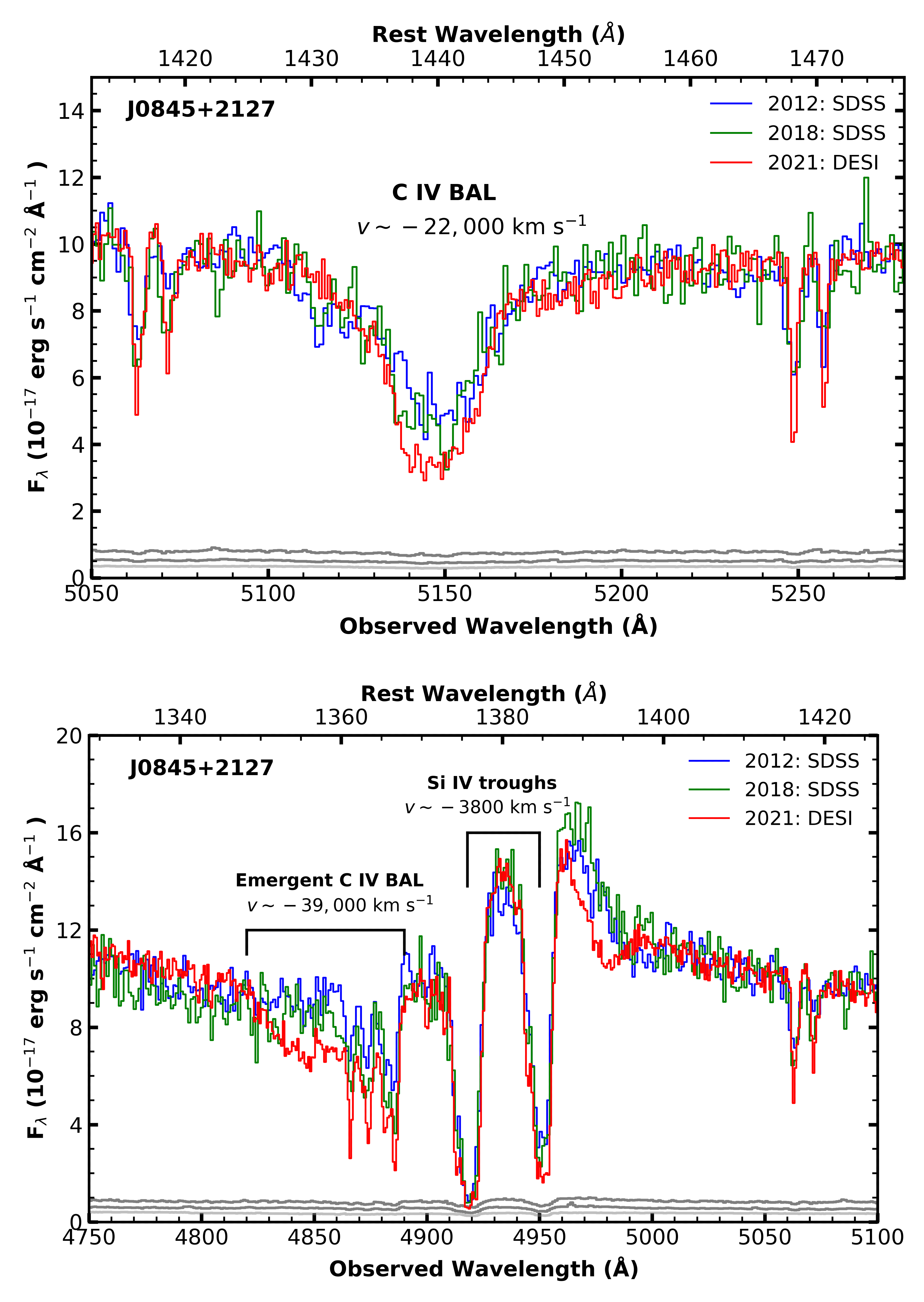}
         \caption{DESI and SDSS spectra of J0845+2127 for three epochs. The continuum flux of the 2018 and 2021 epochs is matched to the 2012 SDSS spectra to highlight the variability in the absorption features. The top panel shows the strengthening of the \ion{C}{IV} BAL with $v \sim-$ 22,000 km s$^{-1}$ over time. The bottom panel shows the emergence of a \ion{C}{IV} BAL with $v \sim-$ 39,000 km s$^{-1}$ between 4820 and 4890 \r{A}. The apparent increase in the depth of the narrow intervening features in both panels is a result of the higher resolution of the DESI spectrum. }
         \label{fig:BALemergence0845}
\end{figure}

To study any possible changes in the absorption features, we scale the 2018 and 2021 epochs to match the continuum flux level of the 2012 epoch. For the main outflowing system (S1), we find no evidence for significant variation in the observed troughs. However as most of the troughs corresponding to S1 are either saturated (e.g., \ion{C}{IV}, \ion{Si}{IV}, \ion{N}{V}) or in the spectral region where the SDSS data has low SNR (e.g., \ion{S}{IV}, \ion{P}{V}), we cannot rule out possible changes. Similarly, we do not find any appreciable variation in the troughs associated with S2. For S3, however, we find that the depth of the \ion{C}{IV} BAL increases gradually from 2012 to 2021 (as shown in top panel of Fig. \ref{fig:BALemergence0845}). The \ion{Si}{IV} 1394 \r{A} trough detected for this system in the 2021 DESI spectrum is not detected in the SDSS spectrum from 2012. In the 2018 spectrum, a weak trough appears, which becomes deeper and wider over $\sim$ 0.9 yrs in the quasar rest frame, leading to the observed trough in the 2021 spectra. A similar behavior is also seen for the \ion{Si}{IV} 1403 \r{A} trough, however, as it is contaminated by a \ion{C}{II} trough of another system, the change is less evident. \par
We also find evidence for emergence of an extremely high velocity BAL system at $v$ $\approx$ $-$39,000 km s$^{-1}$. In the observed wavelength range 4820 \r{A} $\lesssim \lambda_{obs} \lesssim$ 4890 \r{A}, increased absorption is seen in the 2021 spectrum as compared to both 2012 and 2018 spectra (as shown in bottom panel of Fig. \ref{fig:BALemergence0845}). In particular, a new broad absorption feature emerges (seen most clearly for $\lambda_{obs} \lesssim$ 4860 \r{A} where no prior absorption was present) leading to a downwards shift for the narrow absorption features between 4860 and 4890 \r{A}. This broad emergent feature does not align with any expected troughs from all other absorption systems in the spectrum (S1-3) and is thus marked as \ion{C}{IV} corresponding to an extremely high velocity outflow \citep[EHVO; based on the criterion used by][]{hidalgo2020survey}. We also find \ion{N}{V} absorption associated with this system which shows remarkable kinematic correspondence with the absorption profile of \ion{C}{IV}, further confirming our detection of the EHVO. \par
The variation seen in the absorption features in the three epochs could be due to: (a) transverse motion of the outflowing gas across our line of sight \cite[e.g.,][]{moe2009quasar, hall2011implications} and/or (ii) a change in the ionization state of the gas \cite[e.g.,][]{sharma2025physical}. The detected pattern of variation in multiple kinematic components favors the second scenario as suggested by \cite{hamann2011high} and \cite{capellupo2013variability}. This is also supported by the observed change in the continuum flux density between the epochs, which could be correlated with a change in the incident ionizing flux and consequently in the ionization state of the gas.  

\subsection{J1035+3253: Emergence of a BAL component}
\begin{figure}[]
         \centering
         \includegraphics[width=1.00\linewidth]{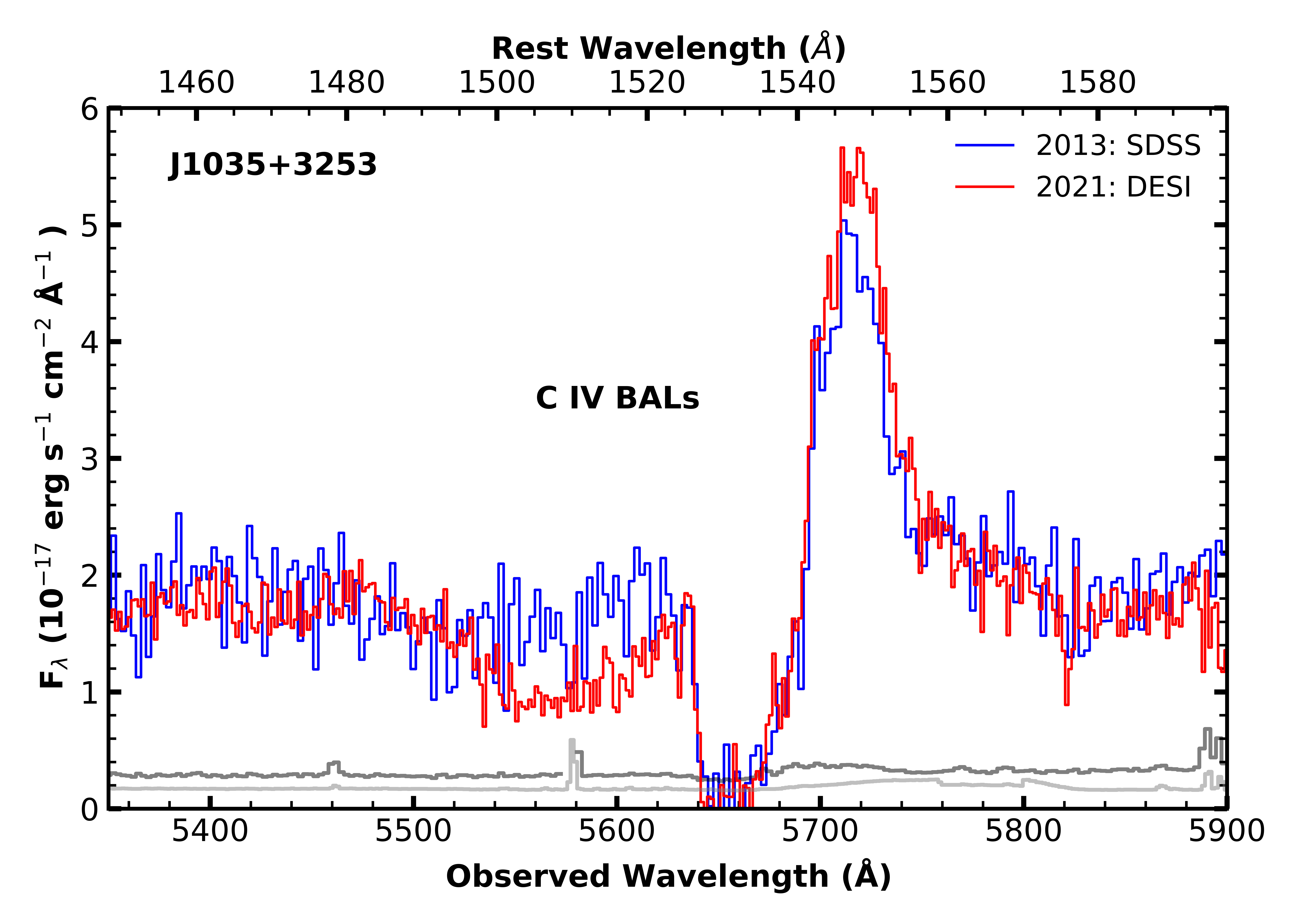}
         \caption{DESI and SDSS spectra of J1035+3253 showing the emergence of a \ion{C}{IV} BAL feature with $v$ $\sim$ $-$8000 km s$^{-1}$.}
         \label{fig:BALemergence1035}
\end{figure}

In the DESI spectrum of J1035+3253, the main outflowing system at $v$ = $-$2570 km s$^{-1}$ (S1) is accompanied by a broad ($\Delta v \approx$ 6000 km s$^{-1}$) high velocity component at $v$ $\sim$ $-$8000 km s$^{-1}$ (S2). This component is only seen in high-ionization species such as \ion{C}{IV}, \ion{O}{VI} and possibly \ion{N}{V} (which is blueshifted to the top of the \ion{Ly}{$\alpha$} emission feature and is thus hard to characterize). The object was also observed as part of SDSS in March 2013 and we compare this spectrum with the DESI spectrum obtained in January 2021 to study any possible variations. We find that the continuum flux decreased by $\sim$ 20\% between the two epochs. After scaling the DESI spectrum to match the continuum flux level of the SDSS spectrum, we find no variation for the detected troughs corresponding to S1. For the high velocity component (S2), however, we find that the \ion{C}{IV} trough is not detected in the SDSS spectra and thus the BAL feature emerges between the two epochs separated by $\sim$ 2 yrs in the quasar rest frame (as shown in Fig. \ref{fig:BALemergence1035}). We are unable to study any potential variation in the \ion{O}{VI} trough associated with S2 due to low SNR of the SDSS spectrum in the wavelength range of interest. The emergence of the \ion{C}{IV} BAL could be due to a change in the ionization structure of the high velocity outflowing gas, which could be associated with the observed change in the continuum flux. However, due to the limited information about the nature of the variability, we cannot discard transverse motion being responsible for the emergence either.  

\section{Discussion} \label{sec:discussion}

\subsection{Possibility of Multiple Ionization Phases}\label{sec:multiphase}

Our photoionization modeling assumes that for a given outflowing component, all the observed ionic species can be modeled with a single ionization phase. This assumption holds well for most of the observed species in the sample, except for \ion{P}{V}, \ion{N}{V} and \ion{O}{VI}. The measured column density of \ion{P}{V} in J0845+2127 and J1426+3151 (see Fig. \ref{fig:nvu}), as well as the obtained lower limits for the column densities of \ion{N}{V} and \ion{O}{VI} in J1021+3209 are underpredicted by the ionization phase that reproduces all other observed species adequately. This provides strong evidence for the existence of another phase with higher ionization material in at least the three above-mentioned outflows. As \cite{arav2020hst} show, this is a common occurrence in outflows that can be studied below $\lambda_{rest} < 1050$ \r{A} allowing for the detection of multiple species with high ionization potential (e.g., \ion{Ne}{VIII}, \ion{Mg}{X} and \ion{Si}{XII}) that can constrain the higher ionization phase. For quasars at cosmic noon, however, DESI is unable to detect troughs from such species and thus we do not have enough constraints for the higher ionization phase. This does not affect our distance determination as the multiple phases are expected to be co-spatial due to their kinematic correspondence. However, as the higher ionization phase is expected to have significantly higher $N_H$, it will carry a much larger mass (see equation \ref{eqn:energetics}). The total energetics of the outflows, therefore, is expected to be higher than the determination based on the single phase solution.

\subsection{Contribution to AGN Feedback}

Theoretical models and simulations have revealed that outflowing winds that carry mechanical energy equivalent to $~\sim$ 5\% of the quasar's luminosity can impart enough energy and momentum to unbind and entrain the cold, dense ISM, which dominates the total gas mass \citep{scannapieco2004quasar}. In the case of `multi-stage' feedback, however, the outflowing winds only need to drive a significant wind in the (much less massive) hot, diffuse ISM. Upon passing through the cold ISM, this will lead to various instabilities that shred/expand it, making it much more susceptible to `secondary' feedback from the incident quasar radiation field \citep{hopkins2010quasar}. In this scenario, the required energy input from the outflowing gas is reduced by at least an order of magnitude to $\sim$ $0.1-0.5$\% of the quasar luminosity. \par
From our sample, the analyzed outflows in at least three objects (J0542-2003, J0845+2127 and J1426+3151) satisfy this criteria (with $\dot{E_k}/L_{Bol}$ $\gtrsim$ $0.1-0.5$\%) for significant AGN feedback to within a factor of 2. For the outflow in J0857+3149, if the actual distance is close to the derived upper limit, it will also lead to a high enough $\dot{E}_k$ for the same. Moreover, as discussed in section \ref{sec:notes}, J0542-2003 and J0845+2127, have multiple higher velocity outflowing components, which will further add to the total mechanical energy output of the outflowing gas in these objects. The combined sample of 15 \ion{S}{IV} outflows at cosmic noon (with distance determination/limits) therefore includes up to 7 outflows ($\sim$ 46\%) that are expected to have a significant impact on their surroundings.

\section{Summary} \label{sec:summary}

In this work, we studied a sample of quasar outflows at cosmic noon observed by DESI. Out of 130 outflowing systems that show \ion{C}{IV} troughs, we confirm the detection of \ion{S}{IV} troughs in 8 of them (through Monte-Carlo simulations), which are subjected to a detailed analysis. The main results are as follows:

\begin{enumerate}
    \item Photoionization modeling of the outflows reveals that the sample spans over 2 orders of magnitude in $N_H$ and 1.5 orders of magnitude in $U_H$, thus probing a wide range of ionization conditions.
    \item By utilizing excited states from three different species (\ion{S}{IV}, \ion{N}{III} and \ion{Si}{II}), we obtained number density solutions for 7 outflows spanning three orders of magnitude.
    \item Most outflows (at least 5 out of 8) are found to be farther than 100 pc from the central source. This includes two outflows that have $R$ > 1 kpc. For two outflows, we could only obtain lower limits on $R$, which places them at distances closer than 100 and 900 pc.
    \item We studied the physical properties of the combined \ion{S}{IV} sample (15 outflows) and found moderate to strong negative correlation between log($R$) and log($|v|$). The correlation of both $v$ and $\dot{M}$ with $L_{Bol}$, however, was found to be weak.
    \item The energetics of the combined \ion{S}{IV} sample were found to be consistent with spatially resolved outflows in a luminosity and redshift matched sample of quasars. We also find that up to 7 out of the 15 analyzed outflows have enough kinetic luminosity to contribute significantly to multi-stage AGN feedback processes.

\end{enumerate}

\begin{acknowledgements}
 We acknowledge support from NSF grant AST 2106249, as well as NASA STScI grants AR-15786, AR-16600, AR-16601 and AR-17556. We thank Dr. Stefano Carniani for letting us use the right-hand panel of Fig. 9 from his \cite{carniani2015ionised} paper, as the base for our Fig. \ref{fig:MassOutflowCarniani}. This research uses services or data provided by the SPectra Analysis and Retrievable Catalog Lab (SPARCL) and the Astro Data Lab, which are both part of the Community Science and Data Center (CSDC) program at NSF National Optical-Infrared Astronomy Research Laboratory. NOIRLab is operated by the Association of Universities for Research in Astronomy (AURA), Inc. under a cooperative agreement with the National Science Foundation.
\end{acknowledgements}

\bibliographystyle{aa}
\bibliography{aanda}

\begin{appendix} 

\section{Spectral Analysis} \label{sec:AppendixA}

\subsection{Identifying \ion{S}{IV} troughs}

Direct identification of the density sensitive \ion{S}{IV} 1063 \r{A} and \ion{S}{IV}* 1073 \r{A} troughs in quasar outflows is hindered by the dense \ion{Ly}{$\alpha$} forest, particularly for high-redshift objects ($z \gtrsim$ 2.5) that make up our sample. However, previous studies have shown that the kinematic profile of the \ion{S}{IV} trough shows good correspondence with that of \ion{Si}{IV} \citep{borguet2012major, borguet2012bal, borguet201210, dunn2012bal, arav2018evidence, xu2018mini, xu2019vlt, dehghanian2025energetic}. Therefore, to aid our identification, we utilize the kinematic profile of the \ion{Si}{IV} trough (\citealp[similar to the methods used by][]{arav2018evidence} and \citealp{xu2019vlt}).
First, we create a template of the \ion{Si}{IV} 1403 \r{A} trough using the Apparent Optical Depth (AOD) model, which relates the optical depth profile with the observed residual intensity as follows:
\begin{equation}
    \tau_{\ion{Si}{IV}}(\lambda) = - \textrm{ln}(I_{\ion{Si}{IV}}(\lambda))
\end{equation}
In cases where the \ion{Si}{IV} 1403 \r{A} trough is contaminated, we use the \ion{Si}{IV} 1394 \r{A} trough. Similarly, when the \ion{Si}{IV} troughs appear heavily saturated, we use the \ion{Al}{III} 1863 \r{A} trough to obtain our template. Having obtained the template, we can create a model for the \ion{S}{IV} trough by shifting the template to the expected wavelength. The model can thus be expressed as:

\begin{equation}
    \tau_{\ion{S}{IV}}(\lambda) = \textit{a} \times  \tau_{\ion{Si}{IV}}(\lambda = \lambda_{\ion{S}{IV} \textrm{, exp.}} ),
\end{equation}
where $a$ is a scale-factor that accounts for difference between the depths of the two troughs and \(\lambda_{\ion{S}{IV} \textrm{, exp.}}\) is the expected wavelength of the \ion{S}{IV} trough, which can be obtained based on the \ion{Si}{IV} trough as \citep[][equation (10)]{arav2018evidence}: 

\begin{equation} \label{expeclambda}
    \lambda_{\ion{S}{IV} \textrm{, exp.}} = \lambda_{\ion{Si}{IV} \textrm{trough}} \frac{1062.66}{1402.77}
\end{equation}

To obtain a model for the \ion{S}{IV} trough, we set the optical depth of the deepest position of the template to 0.2 (which sets the minimum residual intensity for a trough to be detected as $I \approx$ 0.8) and iteratively increase it in steps of 0.01 until any part of the scaled template falls below the maximum flux level allowed by the spectral data (including a 1$\sigma$ noise). We compare our best-fit model with the data by using a modified $\chi^2$ ($\chi^2_{mod}$) defined as:

\begin{equation} \label{chimod}
  \chi^2_{mod} = \frac{1}{(1-\textrm{min(model)})^2} \left[\frac{1}{n-1}\sum_{j}\left(\frac{\textrm{model(\textit{j})-data(\textit{j})}}{\sigma(j)}\right)^2\right] 
\end{equation}
where the sum is over all the points in the template (of which there are $n$) and min(model) refers to the minimum flux predicted by the template (flux corresponding to the deepest point). Within the square brackets, we have the usual definition of reduced-$\chi^2$. The modification we introduced scales the reduced $\chi^2$ such that it is lower for templates that lead to deeper troughs. They are thus preferred by our fitting-algorithm which discourages the misidentification of shallow Ly $\alpha$ forest features as a \ion{S}{IV} trough. \par
To further ensure that our detection of \ion{S}{IV} is robust, we employ Monte Carlo simulations based on the \ion{Si}{IV} template. This is done by randomly selecting 1000 wavelength points in the spectral region surrounding the expected wavelength of the \ion{S}{IV} trough (this is typically 1038-1105 \r{A} in the quasar rest frame). The \ion{Si}{IV} template is shifted to each of these points individually (by replacing 1062.66 in equation \ref{expeclambda} with the randomly chosen test wavelength) and a corresponding best-fit model is obtained. We compare the $\chi^2_{mod}$ for these 1000 models with best-fit model at the expected \ion{S}{IV} wavelength (equation \ref{expeclambda}), and only claim a detection when less than 5\% of the random wavelength models have a better fit to the data then the expected \ion{S}{IV} trough \citep[This procedure is shown visually in Fig. 5 of][]{arav2018evidence}. Based on this selection criteria, we find that in our sample of 130 objects that show troughs from both \ion{C}{IV} and \ion{Si}{IV}, a \ion{S}{IV} trough is detected in eight of them. These eight objects (summarized in Table \ref{table:objectpar}) form the final sample that is subjected to an in-depth analysis in this work. 

\subsection{Column Density Determination}

\begin{figure*}
   \centering
   \resizebox{\hsize}{!}
        {\includegraphics[width=1.00\linewidth]{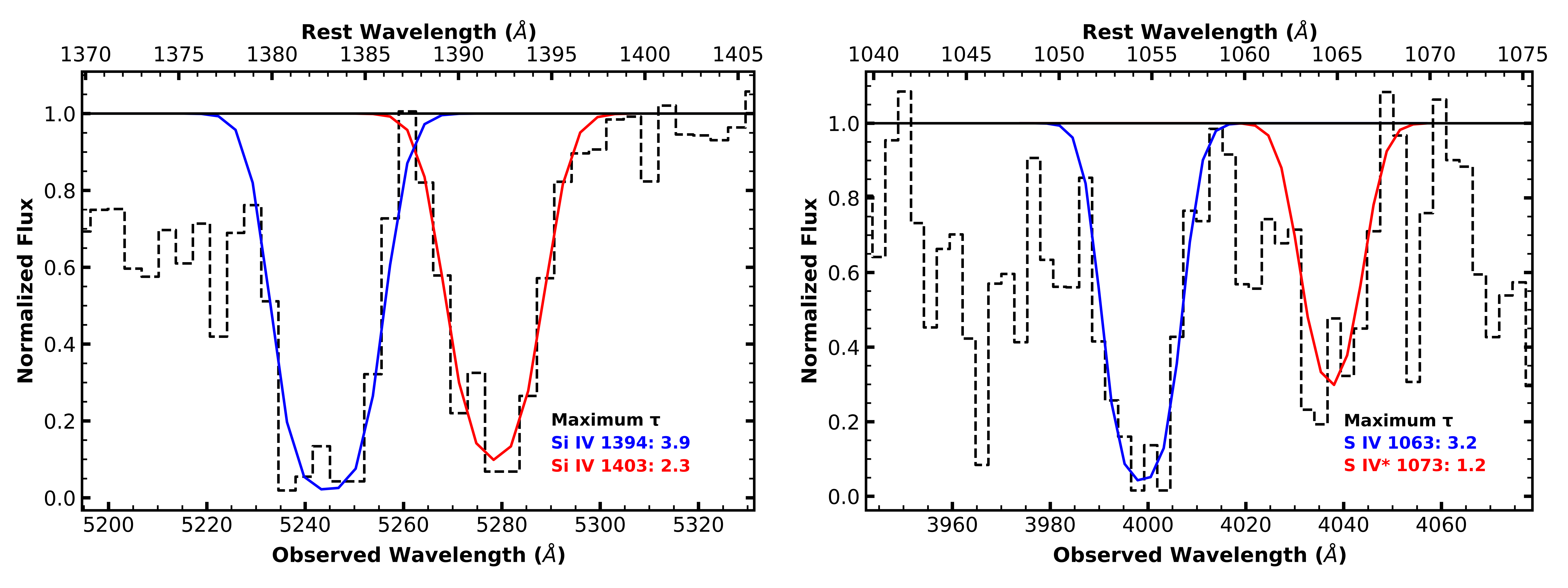}}
      \caption{Example of our template fitting procedure for the absorption troughs in J1426+3151. \ion{Si}{IV} (right panel) and \ion{S}{IV} (left panel) troughs detected in the DESI spectrum are shown with the dashed black histograms. We performed a Gaussian fit to the \ion{Si}{IV} 1403 trough and used its template to fit the others, while allowing for the maximum optical depth to vary. The best-fit models for all four troughs are shown with the solid lines and are accompanied by their resulting best-fit parameter. }
         \label{fig:troughmodel}
   \end{figure*}

For an ionic transition with rest wavelength $\lambda$ (in Angstroms) and oscillator strength $f$, $N_{ion}$ can be obtained as \citep{savage1991analysis}:
\begin{equation}{\label{nion}}
    N_{ion} = \frac{m_e c}{\pi e^2 f \lambda} \int \tau(v) \textrm{ dv} \simeq \frac{3.8 \times 10^{14}}{f \lambda} \int \tau(v) \textrm{ dv} \textrm{\hspace{0.3cm}[cm$^{-2}$]}
\end{equation}
where $m_e$ is the mass of electron, c is the speed of light, $e$ is elementary charge and $\tau(v)$ is the optical depth profile for the transition. To obtain $\tau(v)$, we use the apparent optical depth (AOD) model, which assumes that the central source is completely covered by a homogeneous outflow. This relates the normalized intensity profile ($I(v)$) with $\tau(v)$ as \(I(v) = e^{-\tau(v)}\). Based on this, we model the detected troughs for each outflow component individually. To do so, we first model the \ion{Si}{IV} 1403 \r{A} trough with a single-Gaussian profile for $\tau(v)$ and obtain the best-fit velocity centroid ($v_0$), width ($\sigma$) and depth ($a$) for the trough.  We then create a template with its centroid and width fixed to the obtained best-fit parameters and use it to model all other observed troughs by varying the depth and obtain a best fit model for each using non-linear least squares (as shown in Fig. \ref{fig:troughmodel}). In three of the outflows (J0542-2003, J1021+3209, J1426+3151) we find that the lower ionization species (e.g. \ion{Al}{II}, \ion{Al}{III}, \ion{Si}{II} and \ion{C}{II}) show slight differences in kinematic features (centroid and width) as compared to the template based on the higher ionization \ion{Si}{IV}. Similar difference in trough velocities and width based on ionization potential has been reported previously \citep[e.g.,][]{miller2020hst} and is likely a result of ions with different ionization being formed at varying depths within an outflowing cloud. In these cases, we allow for a different template for the low-ionization species based on the \ion{Al}{III} trough and use that to obtain the best-fit model for the remaining ions. Using equation (\ref{nion}), we then obtain the column densities for all the observed troughs based on their best-fit model. 

The AOD model used for the optical depth profile does not take into account non-black saturation effects in line centers \citep[see section 2.2 in][for a detailed description of non-black saturation]{arav2018evidence} and can thus underestimate the column density derived from troughs that appear saturated. Therefore, in such cases, we consider the column density derived from the AOD model as a lower limit. On the other end, in some cases, we detect a shallow trough ($I\lesssim$ 0.85) at the expected wavelength for a transition. As these can be significantly affected by noise, we only consider them as upper limits. Considering all these effects, we report the final column density measurements/limits for all detected trough for each outflow in Table \ref{table:colden}. The reported error bars include a 1-$\sigma$ error in the best-fit absorption models and a 20\% relative error added in quadrature to account for systematic uncertainties \citep[following][]{xu2018mini}.

\begin{table*}[h]
\setlength{\tabcolsep}{4.3pt}
\renewcommand{\arraystretch}{1.7}
\caption{Ionic Column Densities}
\centering
\begin{tabular}{lccccccccc}
\hline\hline
Ion & $\lambda_{rest}$ & J0542 & J0845 & J0857 & J1021 & J1035 & J1054 & J1229 & J1426 \\ 
\hline
\ion{H}{I} & 1216 & c & $>15.05_{-0.10}$ &$>14.90_{-0.13}$ &$>15.13_{-0.14}$ &$>15.27_{-0.13}$ &$>15.07_{-0.11}$ &$>15.15_{-0.15}$ &$>15.49_{-0.13}$\\
\ion{C}{II} & 1335 & $14.57^{+0.11}_{-0.15}$ & - & $14.64^{+0.10}_{-0.13}$ &$14.12^{+0.14}_{-0.20}$ & $14.59^{+0.14}_{-0.22}$ & - & $13.57^{+0.08}_{-0.10}$ & $14.95^{+0.11}_{-0.15}$\\
\ion{C}{II*} & 1336 & $14.72^{+0.11}_{-0.16}$ & - & $15.09^{+0.13}_{-0.19}$ & $14.35^{+0.10}_{-0.13}$ & b & - & $14.03^{+0.13}_{-0.19}$ & b\\
\ion{C}{IV} & 1551 & $>15.36_{-0.10}$ & $>15.60_{-0.10}$ &$>15.75_{-0.13}$ &$>15.45_{-0.11}$ &$>15.89_{-0.15}$ &$>15.51_{-0.14}$ &$>15.32_{-0.13}$ &$>16.10_{-0.20}$\\
\ion{N}{III} & 990 & - & - & - & - & - & - & $15.35^{+0.11}_{-0.14}$ & -\\
\ion{N}{III*} & 992 & - & - & - & - & - & - & $15.57^{+0.11}_{-0.15}$ & -\\
\ion{N}{V} & 1243 & $>15.29_{-0.10}$ & $>15.78_{-0.10}$ &$>15.83_{-0.11}$ &$>15.74_{-0.12}$ &$>15.99_{-0.15}$ &$>15.65_{-0.14}$ &$>15.69_{-0.13}$ &$>16.06_{-0.12}$\\
\ion{O}{VI} & 1038 & $>15.51_{-0.10}$ & $>16.00_{-0.20}$ &$>16.06_{-0.24}$ &$>15.73_{-0.10}$ &$>15.95_{-0.31}$ &$>15.76_{-0.25}$ &$>16.04_{-0.28}$ &$>16.22_{-0.16}$\\
\ion{Al}{II} & 1671 &$12.74^{+0.09}_{-0.12}$ & - &$13.03^{+0.09}_{-0.11}$ & - & - & - & - & -\\
\ion{Al}{III} & 1863 & $13.99^{+0.09}_{-0.12}$ & $13.29^{+0.11}_{-0.15}$ &$14.50^{+0.08}_{-0.10}$ & $13.04^{+0.10}_{-0.14}$ & $14.47^{+0.10}_{-0.13}$ & $15.44^{+0.10}_{-0.13}$ & - & $14.36^{+0.10}_{-0.13}$\\
\ion{Si}{II} & 1260 & $13.34^{+0.09}_{-0.12}$ & - & - &$13.07^{+0.09}_{-0.11}$ & - & - & - & $13.55^{+0.10}_{-0.13}$\\
\ion{Si}{II*} & 1265 & $13.53^{+0.09}_{-0.11}$ & - & - &$12.40^{+0.17}_{-0.27}$ & - & - & - & $13.78^{+0.10}_{-0.13}$\\
\ion{Si}{III} & 1207 & c & c & $>14.06_{-0.13}$ &  $>14.00_{-0.11}$ & c & c & c & $>14.73_{-0.13}$ \\
\ion{Si}{IV} & 1403 &$>14.79_{-0.10}$ & $>15.14_{-0.13}$  & $>15.22_{-0.12}$ &$>14.64_{-0.15}$ &$>15.41_{-0.14}$ &$>14.65_{-0.13}$ &$>14.36_{-0.13}$ &$>15.34_{-0.14}$\\
\ion{P}{V} & 1128 & - & $14.67^{+0.09}_{-0.11}$ & - & - & - & - & - & $15.43^{+0.12}_{-0.16}$\\
\ion{S}{IV} & 1063 &$15.65^{+0.09}_{-0.11}$ & $15.54^{+0.09}_{-0.12}$ & $15.75^{+0.09}_{-0.11}$ & $15.23^{+0.09}_{-0.12}$ & $>16.38_{-0.19}$ & $15.36^{+0.11}_{-0.15}$ & $15.23^{+0.11}_{-0.15}$ & $16.31^{+0.09}_{-0.11}$\\
\ion{S}{IV*} & 1073 &$<14.79^{+0.10}$ & $15.23^{+0.08}_{-0.10}$ & $>15.95_{-0.10}$ & $<14.73^{+0.14}$ & $>16.09_{-0.15}$ & $15.25^{+0.11}_{-0.15}$ & $15.25^{+0.09}_{-0.11}$ & $15.96^{+0.10}_{-0.13}$\\

\hline
\end{tabular}
\tablefoot{The ions marked with a dash (-) were not identified in the spectra of the corresponding object. \ion{C}{II*} troughs marked with a (b) were blended with \ion{C}{II} and thus their column density is included in the reported \ion{C}{II} values. \ion{Si}{III} troughs marked with a (c) were contaminated and thus reliable column density limits could not be obtained for them.}
\label{table:colden}
\end{table*}

\section{Photoionization Modeling} \label{sec:AppendixB}

We model the outflowing cloud as a plane-parallel slab of constant hydrogen number density ($n_H$) and solar abundance, which is irradiated upon by the flux from the central source. This cloud can then be characterized by its total hydrogen column density ($N_H$) and ionization parameter ($U_H$) which is related to the rate of ionizing photons emitted by the source ($Q_H$) by:
\begin{equation*} 
    U_H = \frac{Q_H}{4\pi R^2 n_H c}
\end{equation*} 
Here $R$ is the distance between the emission source and the observed outflow component and $c$ is the speed of light. $Q_H$ is determined by the choice of the spectral energy distribution (SED) that is incident upon the outflow and the redshift of the object. For all objects in our sample (except J0845+2127, see below), we adopt the UV-soft SED of \cite{dunn2010quasar} and consider the effect of using different SEDs later in the section.

\subsection{Determining the best-fit solution}

For each object, we create a grid of models in the ($N_{H}, U_{H}$) phase space predicting the column densities of all the relevant ions at each point. These prediction are then compared to the measured ionic column densities ($N_{ion}$) to constrain the phase space for each individual ionic species. The overall best-fit solution for the outflow is then determined by minimizing the $\chi^2$ between the model predictions and the actual measurements/limits. \par 

\subsection{Dependence on SED} 

The choice of the incident SED is an important factor in determining the ionization and thermal structure of the outflowing gas. The UV-soft SED used primarily in our analysis was constructed by \cite{dunn2010quasar} based on models of accretion disk emission and improved measurements of the UV and X-ray continuum for luminous quasars (see their section 4.2 for a detailed description). To understand how sensitive our photoionization solution is to this choice, we consider the effect of two other SEDs: (i) The SED predicted by \cite{mathews1987heats} (hereafter MF87) and (ii) The best empirically determined SED in the extreme-UV, constructed for the quasar HE 0238-1904 by \cite{arav2013quasar} (hereafter HE0238). \par

For all objects in our sample (except J0845+2127), we find that choosing either MF87 or HE0238 SED for our photoionization modeling changes the best-fit solutions for both $N_H$ and $U_H)$ (determined using the UV-soft SED) by less than 0.2 dex. In the case of J0845+2127, however, the effect is much more pronounced. This is because its photoionization solution is relatively poorly constrained as column density measurements could only be obtained from two ionic species (\ion{Al}{III} and \ion{S}{IV}). For the UV-Soft and MF87 SEDs, the contours corresponding to these species in the ($N_H,U_H$) phase-space plot overlap for a large range of parameters. The $1-\sigma$ contours for the best-fit solution, therefore, effectively cover the entire range, making the solution unreliable. For the HE0238 SED, the overlap is over a much smaller range in parameters and thus the best-fit solution and its $1-\sigma$ contour are well constrained. We thus use this solution for the rest of the analysis, while noting its strong dependence on the SED.  

\end{appendix}

\end{document}